\documentclass[journal,article,submit,moreauthors,pdftex,10pt,a4paper]{mdpi}
\graphicspath{{./figs/}} 

\firstpage{1} 
\articlenumber{x}
\doinum{10.3390/------}
\pubvolume{xx}
\pubyear{2017}
\copyrightyear{2017}
\externaleditor{Academic Editor: name}
\history{Received: date; Accepted: date; Published: date}

\pdfoutput=1

\Title{Centrality dependent L\'evy-stable two-pion Bose-Einstein correlations in $\sqrt{s_{\textmd{NN}}}=200$ GeV Au+Au collisions at the PHENIX experiment}

\Author{S\'andor L\"ok\"os for PHENIX Collaboration$^{1,2,*,\dagger}$}

\AuthorNames{S\'andor L\"ok\"os}

\address{%
$^{1}$ \quad Eszterh\'azy K\'aroly University, H-3200 Gy\"ongy\"os M\'atrai \'ut 36, Hungary \\
$^{2}$ \quad Eötvös University, H-1117 Budapest, Pázmány P. s. 1/A, Hungary\\
}
\corres{Correspondence: lokos@caesar.elte.hu}

\abstract{Investigation of femtoscopic correlation functions in relativistic heavy ion reactions is an important tool to access the space-time structure of particle production in the strongly interacting Quark Gluon Plasma (sQGP). The shape of the source thus the shape of the correlation functions is often assumed to be Gaussian, but experimental results found evidence for heavy tails in the source distribution of pions. Recent analysis revealed that the statistically correct assumption could be the so-called L\'evy distribution. The detailed investigation of correlation functions in various systems may shed light on the location of the critical endpoint on QCD phase diagram. It could also reveal if there is partially coherent pion production or could indicate the possible in-medium mass modification of the $\eta'$ meson due to the (partial) restoration of the $U_{\rm A}(1)$ axial symmetry. These phenomena could depend on the system size and on the collision energy. A detailed centrality dependent analysis could explore the multiplicity dependencies of the L\'evy parameters thus the critical and thermodynamical properties of the sQGP and could give information about the above mentioned processes. In this paper we present the status of the centrality dependent measurements of two-pion L\'evy Bose-Einstein correlation functions $\sqrt{s_{\textmd{NN}}}=200$ GeV Au+Au collisions at PHENIX.}


\begin{document}

\section{Introduction}

Intensity correlation measurements represent a widely used technique in high energy physics. The technique was discovered in astrophysics by R. Hanbury Brown and Q. R. Twiss in correlation measurements, that were performed in radio and optical astronomy to measure the angular diameters of stars \cite{HBT_orig}. Hanbury Brown and Twiss are considered the experimental founders of the HBT effect. Independently, the intensity correlations of identical pions were observed in proton-antiproton annihilation. These correlations were explained by G. Goldhaber, S. Goldhaber, W-Y. Lee and A. Pais \cite{PhysRev.120.300} on the basis of the Bose-Einstein symmetrization of the wave-function of identical pion pairs. 

Because the Bose-Einstein correlations are related to the Fourier transform of the source, correlation measurements can provide an insight into the shape and the dynamics of the source. The shape of the source is usually assumed to be Gaussian universally across collision energy, particle type and centrality. PHENIX found evidence for and power-law tails in the source distribution for pions \cite{Adler:2006as}. To understand the underlying processes better one has to go beyond the Gaussian approximation to a more general distribution: the L\'evy distribution \cite{Csorgo:2003uv, Metzler:1999zz, Csanad:2007fr}.

L\'evy-type of analyses were discussed before in Refs. \cite{Csanad:2005nr, Kincses:2016jsr}. Furthermore, a detailed analysis 0-30\% Au+Au data at $\sqrt{s_{\rm NN}}=200$ GeV has been performed, where further details of Lévy femtoscopy are also given \cite{Adare:2017vig, MateBGL}. In this paper we present the L\'evy-type of correlation functions measured by the PHENIX experiment at various centralities at 200 GeV.

\section{PHENIX experiment}

\begin{figure}[h!]
\centering
\includegraphics[width=0.95\textwidth]{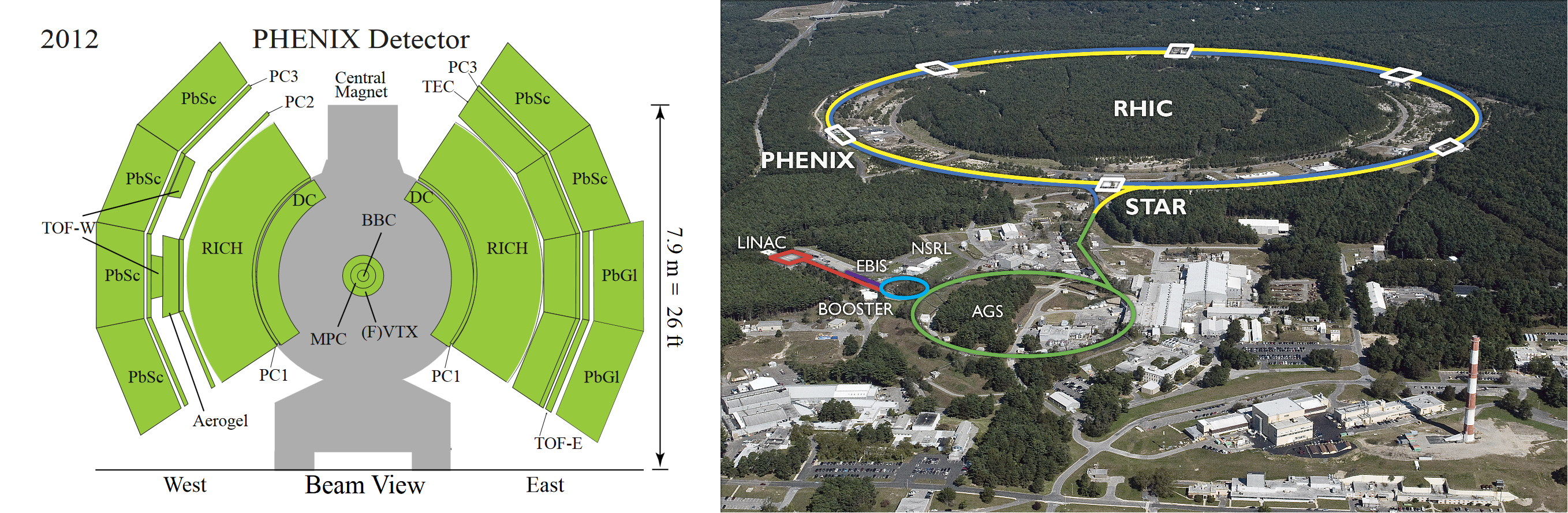}
\caption{\label{fig:phenix_exp} A schematic beam view drawing of the detector system of the accelerator complex at the PHENIX experiment and the RHIC facilities from a bird view.}
\end{figure}

The PHENIX experiment at RHIC was designed to study various type of particles produced in heavy ion collisions. A schematic beam view drawing of the detector system and a bird view photo can be seen in Figure \ref{fig:phenix_exp}. An exhaustive overview of the experiment can be found in Ref. \cite{Adcox:2003zm}, here we restrict ourselves to mention the detectors that played role in the present analysis.

This analysis uses the beam-beam counters (BBC) for event characterization. Its two arms cover 2$\pi$ in azimuth angle. The charge sum in both BBC arms is used to determine the centrality of the event. The tracking and the determination of the longitudinal position of the vertex were done by Drift Chamber (DC) and Pad Chambers (PC1). The charged pion identification was done by lead scintillator Electromagnetic Calorimeter (PbSc) as well as the high resolution time-of-flight detectors (TOF East and TOF West). We identified charged pions in $~ 0.2$ GeV/c $\leq p_{\rm T} \leq 0.85 $ GeV/c transverse momentum range.

\section{Two-particle correlation functions and the L\'evy distribution}
\label{sec:levy}

The detailed theoretical description of the basis of the L\'evy-type of analysis can be found in Ref.~\cite{Adare:2017vig, MateBGL}. Here we just give a short introduction to the topic. In general, the two-particle correlation function is defined as

\begin{equation}
C_2(p_1,p_2)=\frac{N_2(p_1,p_2)}{N_1(p_1)N_1(p_2)},
\label{eq:c2}
\end{equation}
where $N_k$ is the $k$-particle invariant momentum distribution function. With the $S(x,p)$ source function \cite{Yano:1978gk}, the following formula can be obtained for the two-particle correlation function

\begin{equation}
C_2^{(0)}(q,K)=1+\frac{|\tilde{S}(q,K)|^2}{|\tilde{S}(0,K)|^2},
\label{eq:c2_0}
\end{equation}
where the four-momentum difference of the pair $q=p_1-p_2$ and the average momentum of the pair $K=(p_1+p_2)/2$ were introduced as new variables. The $^{(0)}$ index indicates that none of the final state effects were taken into account. Without Coulomb interaction in the final state the correlation function takes the value of 2 at zero relative momentum. However, the experimental limits do not allow us to resolve the relative momentum below 4-8 MeV/c. Hence it is usually observed that the experimentally measured correlation functions, when extrapolated to zero relative momentum, yield a value different from 2. This can be quantified in the following form:

\begin{equation}
\lambda(K) = \lim_{q\rightarrow 0} C_2(q,K)-1 \neq 1.
\label{eq:3}
\end{equation}
This $\lambda(K)$ is called the intercept parameter or the strength of the correlation function. This observation can be interpreted easily with the core-halo model. In this model of the Bose-Einstein correlations \cite{Bolz:1992hc, Csorgo:1994in} the source is treated as a composite one with two parts: a core part which is a hydrodynamically behaving, fireball-type source and a halo part, which contains long-lived resonances.

\begin{figure}[h!]
\centering
\includegraphics[width=0.55\textwidth]{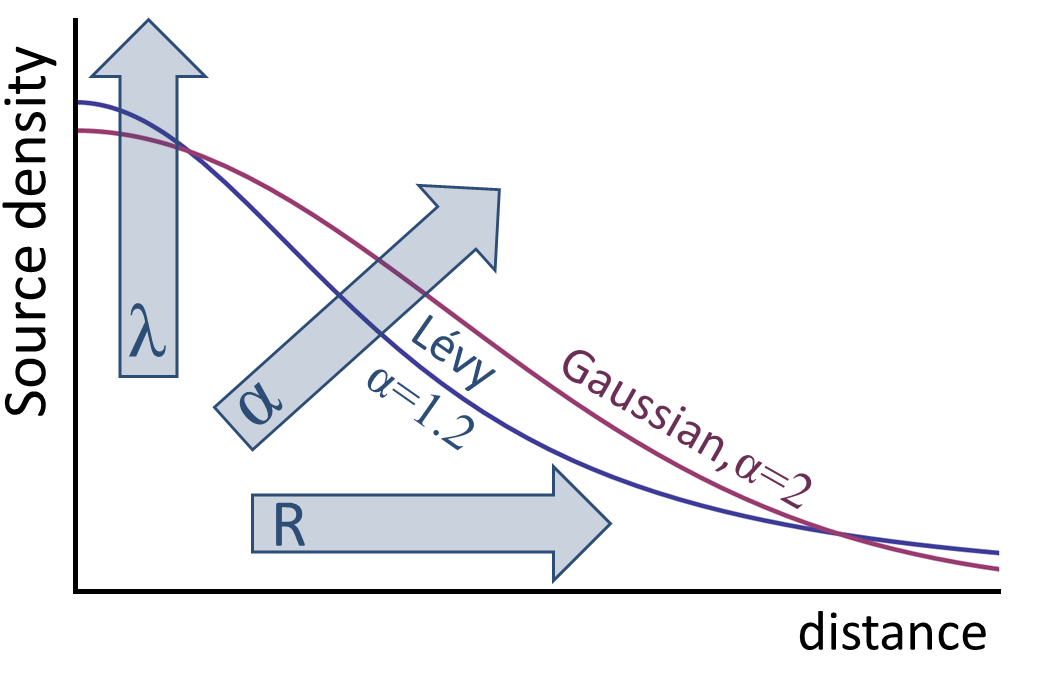}
\caption{\label{fig:Levy_example} Comparison of the Gaussian and the L\'evy distribution with a given $\alpha$, $R$ and $\lambda$. It can be seen that the L\'evy distribution has a power-law-like tail.}
\end{figure}

Several long-lived resonances with a decay width smaller than the resolvable momentum difference (e.g. $\eta, \eta', K_S^0$ or in an appropriate experiment $\omega$) decay to pions, which contribute to the halo region. The pions coming from the core do not correlate with the pions coming from the decays of the resonances of the halo part, but both type of pions will be detected. Therefore the long-lived resonances could reduce the correlation strength in a momentum range given by the decay kinematics. So the correlation function of Eq. \eqref{eq:3} is modified into the

\begin{equation}
C_2(q,K)=1 + \left(\frac{N_c(K)}{N_c(K)+N_h(K)} \right)^2 \frac{|\tilde{S}(q,K)|^2}{|\tilde{S}(0,K)|^2} = 1 + \lambda \frac{|\tilde{S}(q,K)|^2}{|\tilde{S}(0,K)|^2}
\end{equation}
form. If there is no halo part and the source is chaotic and fully thermal then $\lambda=1$. If $\lambda < 1$ it is the sign of the presence of the decay mesons. For more detailed calculations, discussions and conclusions see Ref. \cite{Csorgo:1999sj, Vertesi:2009wf, Csorgo:2010hj}.
\\
\linebreak
Now let us discuss the shape of the correlation functions. In an expandig system, the mean free path is increasing as the system expands. In such case a generalized form of the central limit theorem leads to the appearance of the L\'evy distribution as source function \cite{Csorgo:2003uv, Metzler:1999zz, Csanad:2007fr}. The one-dimensional, symmetric L\'evy distribution is defined by a Fourier transform:

\begin{equation}
\mathcal{L}(\alpha,R,\textbf{r})=\frac{1}{(2\pi)^3}\int d^3 \textbf{q} e^{i\textbf{qr}} e^{-\frac{1}{2} |\textbf{q}R|^\alpha}.
\end{equation}
Here $\alpha$ is called the L\'evy-index of stability or the shape parameter, while $R$ is the L\'evy scale parameter. In the case of $\alpha=2$ the Gaussian case is restored, the case of $\alpha=1$ corresponds to the Cauchy distribution. From the L\'evy-type of source the following correlation function can be obtained:

\begin{equation}
C_2^{(0)}(q,K)=1+\lambda(K) e^{-(R(K)q)^{\alpha(K)}}.
\end{equation}
The effect of the parameters can be seen on Figure \ref{fig:Levy_example}. Previous measurements of two-pion Bose-Einstein correlation functions in Au+Au collisions \cite{Csanad:2005nr} indicate the presence of a long-range power-law-like component which could be understood in terms of L\'evy distributions.
The $\alpha$ shape parameter or L\'evy-index indicates the deviation from the Gaussian case. Moreover, it could be associated to one of the critical exponents namely to the critical exponent of the spatial correlation in 3 dimension \cite{Csorgo:2003uv}. Because of this possible relation between the two exponents, the measurements of $\alpha$ could help to locate the critical point on the phase diagram of the QCD if there is any \cite{Csorgo:2003uv, Csanad:2007fr, Csorgo:2005it}. The critical exponent of the spatial correlation could vary near to a critical point. If $\alpha$ could be associated to this critical exponent, it should show changes near to the critical point. This association of the two exponent gives us the motivation to preform precise HBT measurements in various systems with different sizes or created with different energies.

In a hot, dense matter like the sQGP the $U_{\rm A}(1)$ symmetry might be restored. If it is, then the $\eta'$ meson should have smaller mass than with broken $U_{\rm A}(1)$ which means enhanced production of the $\eta'$ meson. This particle could decay into five pions and because of the dynamics the pions will have ${\color{red}p_{\rm T}} \approx 150-200$ MeV/c transverse momentum. These pions contribute to the halo part of the source which means that the correlation strength will be supressed by the effect around these $p_{\rm T}$ range. So if we measured that the $\lambda$ has a decreasing trend at lower $p_{\rm T}$ range then it might be caused by the (partially) restored $U_{\rm A}(1)$ axial symmetry. A detailed explanation can be found in Ref. \cite{Vertesi:2009wf} and a L\'evy-type analysis of Au+Au collisions at 200 GeV in 0-30\% centrality can be found in Ref. \cite{Adare:2017vig}.
\\
\linebreak
There are other possible explanation for the non-constant behavior of the $\lambda(m_{\rm T})$. Partial coherence could effect the $\lambda(m_{\rm T})$ but the investigation of this phenomena requires 3-particle correlation measurements \cite{Csorgo:1999sj}. Model which take into account the compositeness of the pion and its interactions exhibits similar behavior of the intercept parameter \cite{Gavrilik:2014pxa}.
\\
\linebreak
Because charged pions are measured, the Coulomb interaction in the final state has an effect that should be taken into account. We utilized a generalization of the Sinyukov method detailed in Ref. \cite{Sinyukov:1998fc} similarly to Ref. \cite{Adare:2017vig} to handle this final state effect.

\section{Results of the centrality dependent analysis}

We studied the L\'evy-parameters in 10\% wide centrality bins in 0-60\% range and in 18 transverse mass ($m_{\rm T}$) bins at $\sqrt{s_{\rm NN}}$ = 200 GeV collision energy in Au+Au system. We determine the $m_{\rm T}$ and $N_{\rm part}$ dependencies of the L\'evy parameters. The transverse mass is defined with $m_{\rm T} = \sqrt{m^2 + p_{\rm T}^2}$.

\subsection{The L\'evy shape parameter $\alpha$}

\begin{figure}[h!]
\centering
\includegraphics[width=0.45\textwidth]{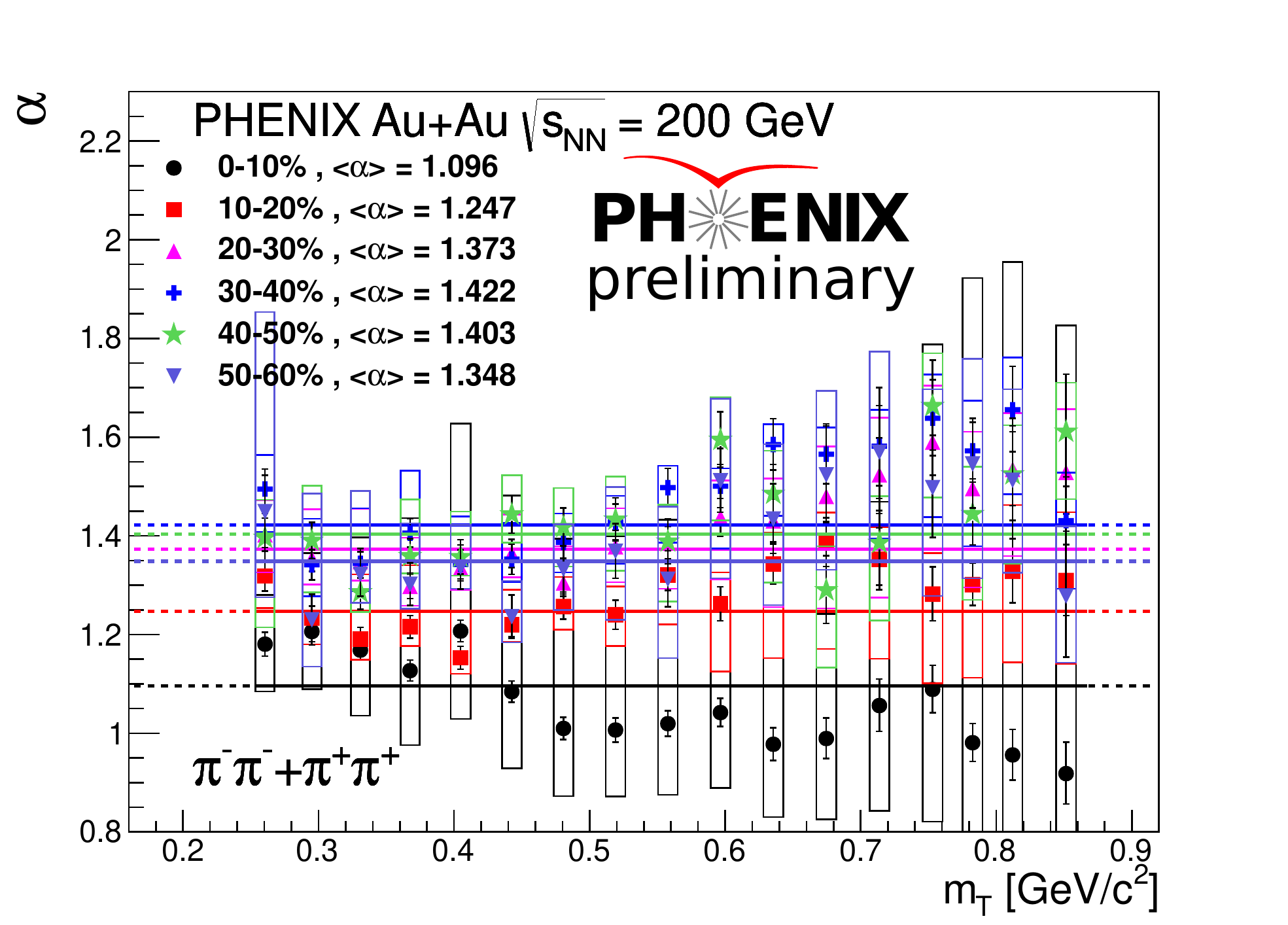}
\includegraphics[width=0.45\textwidth]{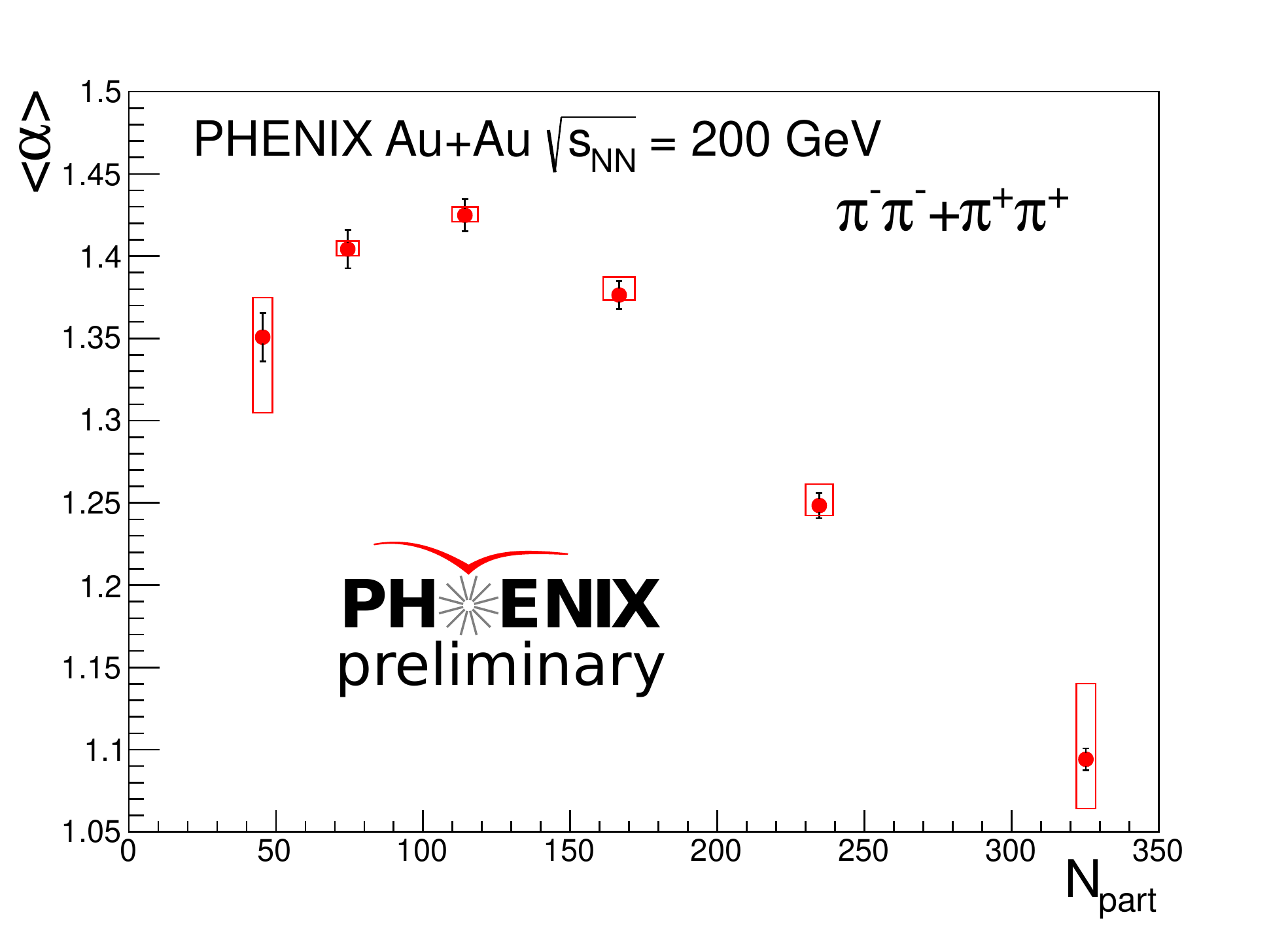}
\caption{\label{fig:alpha} The $\alpha$ L\'evy stability parameter as a function of $m_{\rm T}$ (left). The $\langle \alpha \rangle$ as a function of $N_{\rm part}$ (right). The statistical errors are shown with errorbars and the systematical uncertainties boxes.}
\end{figure}

First, we measured the $\alpha(m_{\rm T})$ dependencies in the above mentioned centrality ranges. The results can be seen in Figure \ref{fig:alpha}. On the left, it can be observed that the value of $\alpha$ does not depend strongly on $m_{\rm T}$, compared to the systematic uncertainties. Hence the $m_{\rm T}$ average $\langle \alpha \rangle$ is meaningful and describes well the exponent for each centrality bin, we could perform fix $\alpha = \langle \alpha \rangle$ fits.
\\
\linebreak
If we study the centrality dependence of $\alpha(m_{\rm T})$ points or the averages it turns out that there is a clear non-monotonic behavior as a function of $N_{\rm part}$ (as shown in the right panel of Figure \ref{fig:alpha}.). We observe that all measured $\alpha$ values are far from the Gaussian limit of $\alpha=2$.

\subsection{The L\'evy scale parameter $R$}

\begin{figure}
\centering
\includegraphics[width=0.45\textwidth]{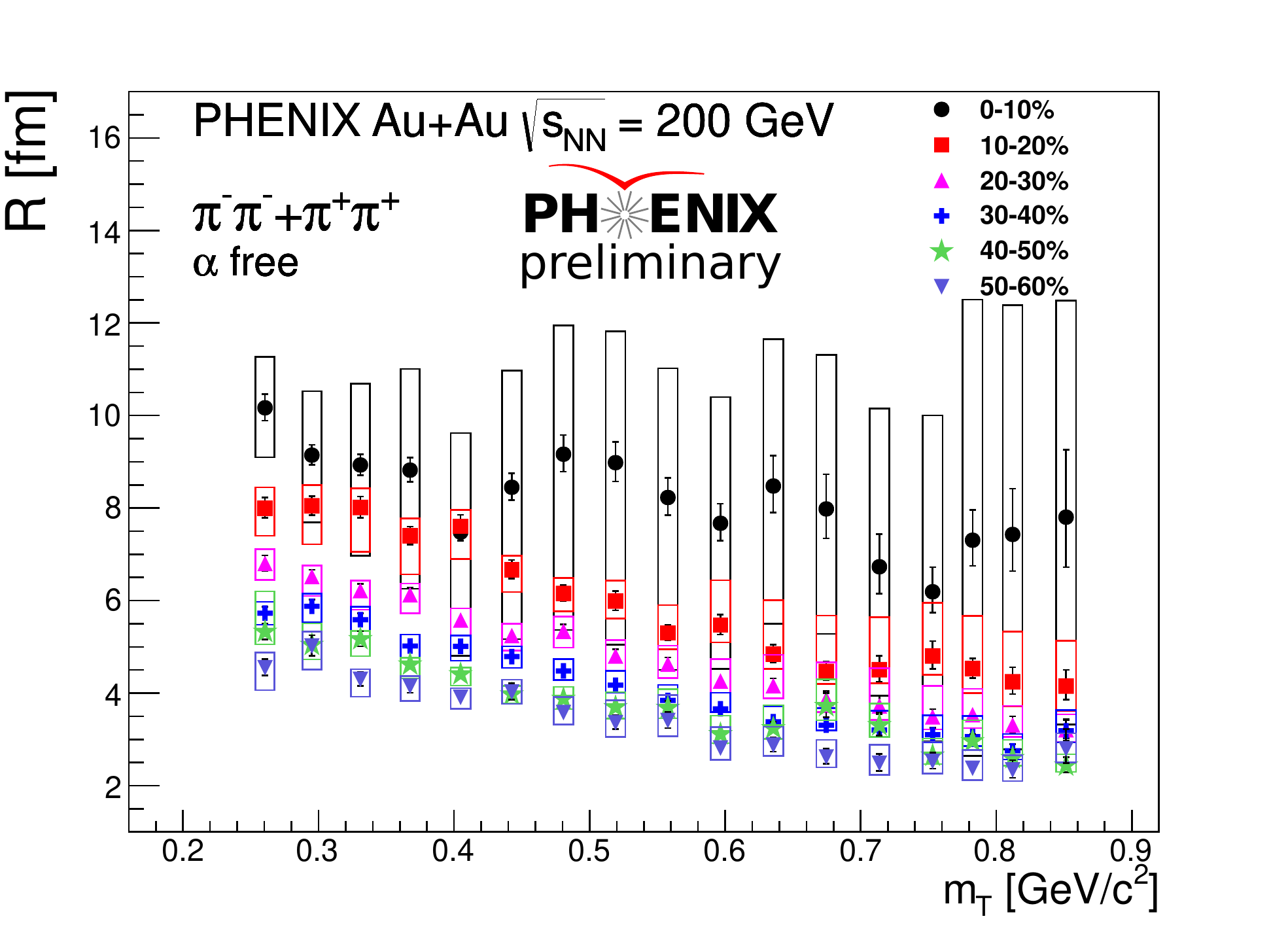}
\includegraphics[width=0.45\textwidth]{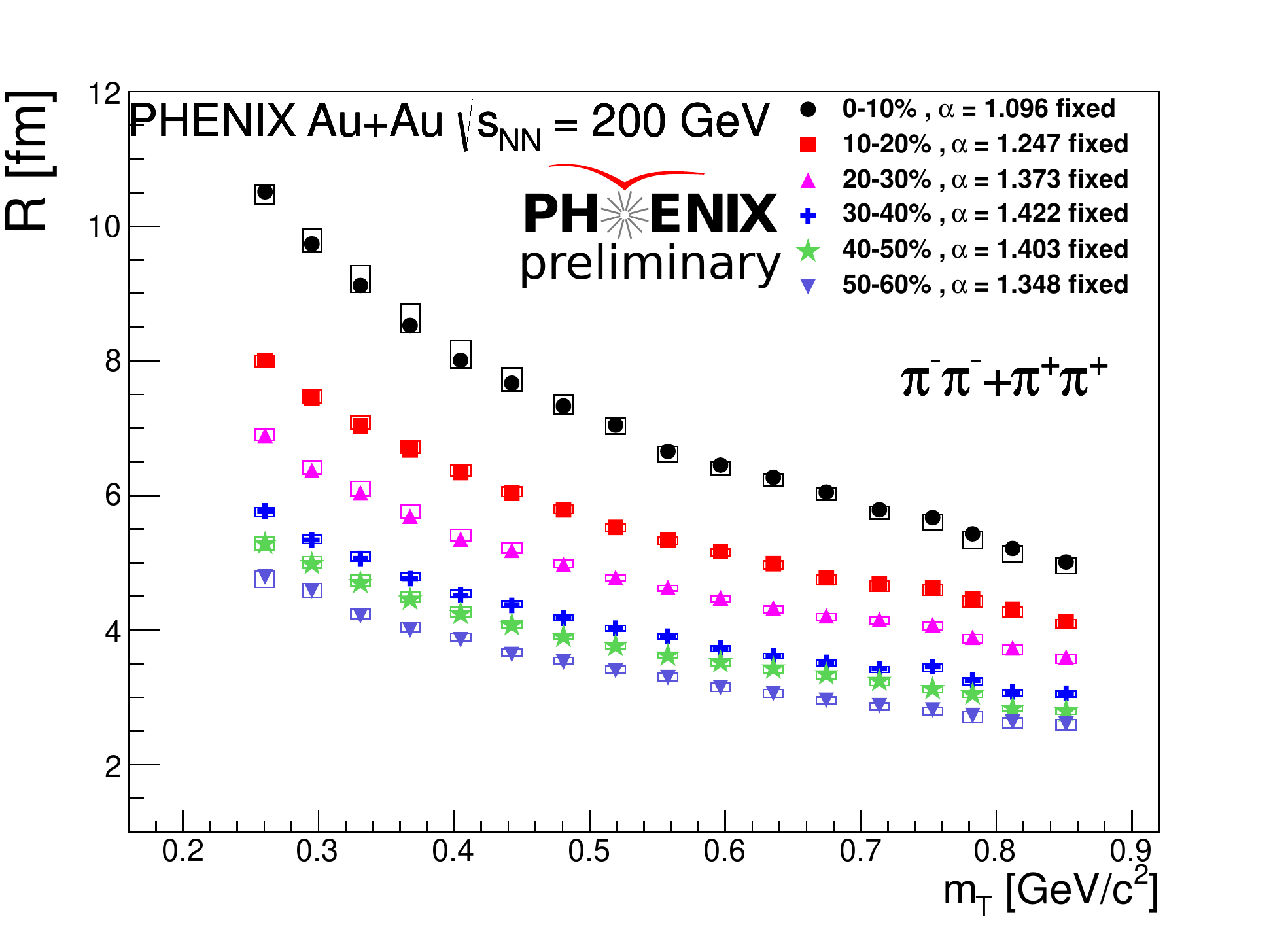}
\caption{\label{fig:R} The L\'evy scale parameter as a function of $m_{\rm T}$ with free $\alpha$ (left) and in the case when $\alpha=\langle \alpha \rangle$ is fixed during the fit (right). The statistical errors are shown with errorbars and the systematical uncertainties boxes.}
\end{figure}

\begin{figure}
\centering
\includegraphics[width=0.45\textwidth]{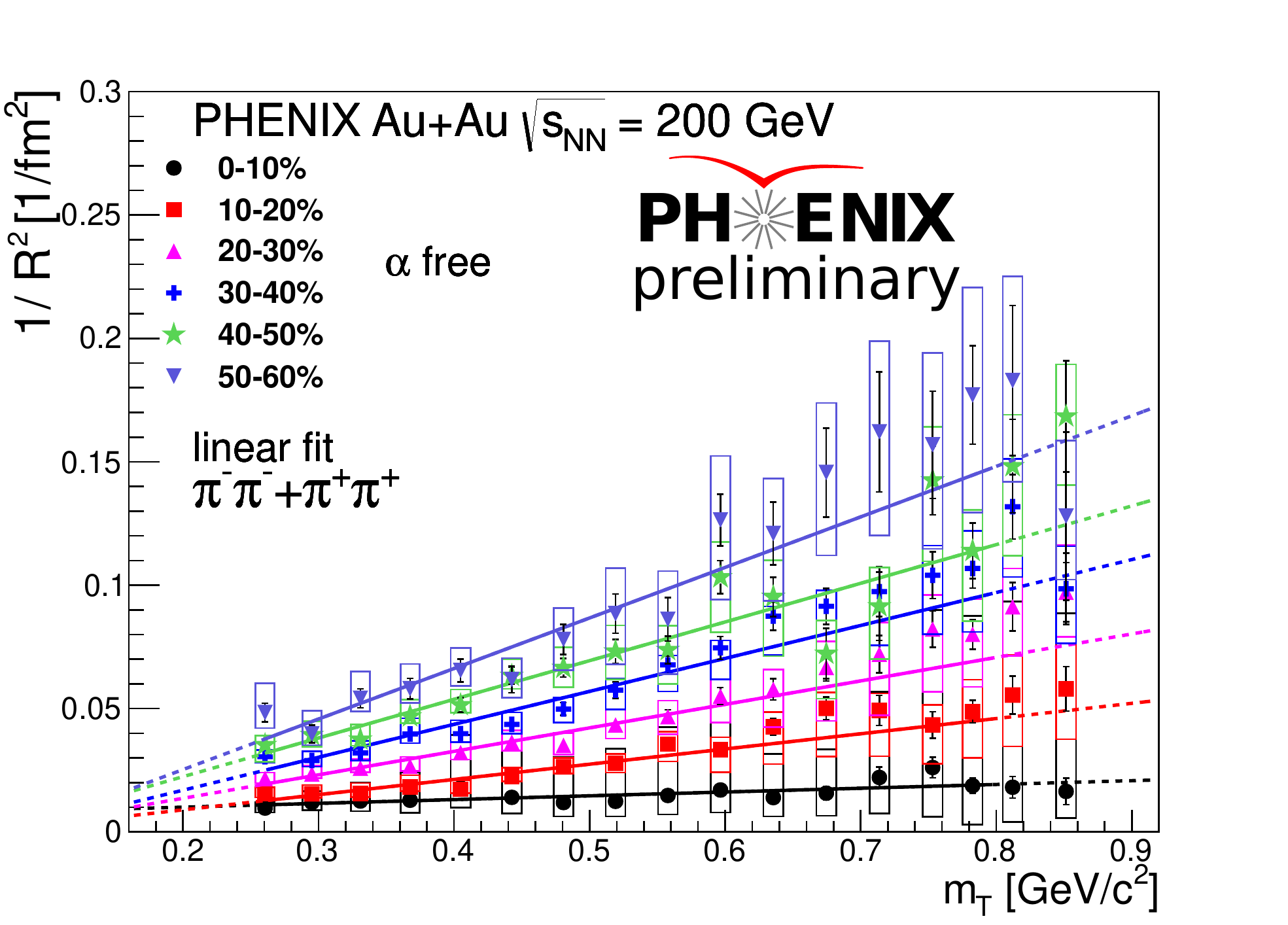}
\includegraphics[width=0.45\textwidth]{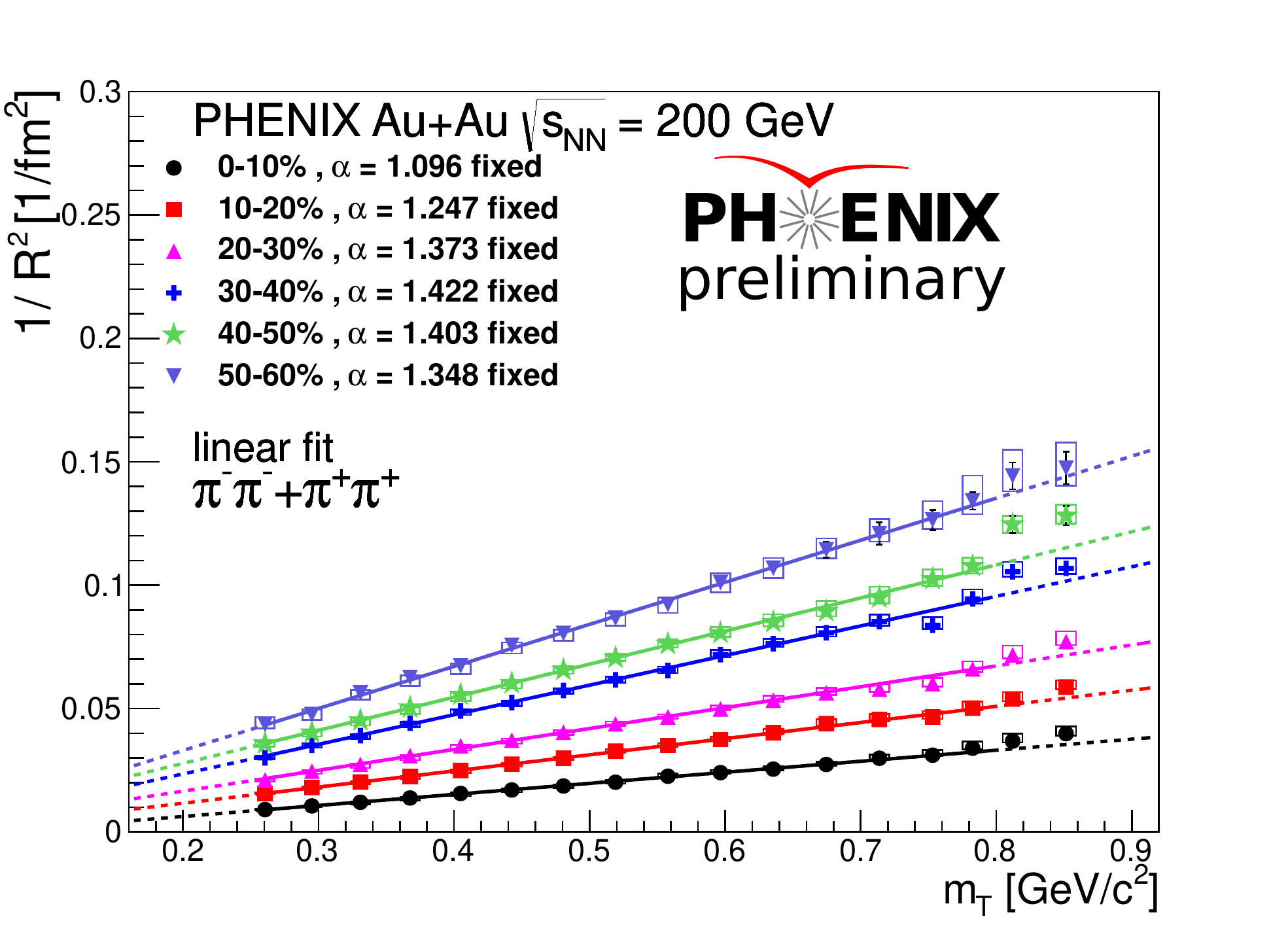}
\caption{\label{fig:1overR2} The hydro-type behavior of the L\'evy scale parameter as a function of $m_{\rm T}$ with free $\alpha$ (left) and in the case when $\alpha=\langle \alpha \rangle$ is fixed during the fit (right). The statistical errors are shown with errorbars and the systematical uncertainties boxes.}
\end{figure}

In a L\'evy case, the source has no second moment, hence the L\'evy scale is not equivalent to the Gaussian radius or RMS value. Nevertheless, the L\'evy scale exhibits similar features, for example a decreasing trend as a function of $m_{\rm T}$ as it can be seen in Figure \ref{fig:R}. in case of free and fix $\alpha$. It turns out that the $1/R^2 \propto a + b \cdot m_{\rm T}$ kind of proportionality (deduced from hydrodynamical calculations \cite{Makhlin1988, Csorgo:1995bi, Chapman:1994yv}) remains valid in the L\'evy case too. This is visualized in Figure \ref{fig:1overR2}., where we show $1/R^2$ vs ${m_{\rm T}}$, also in case of free and fix $\alpha$. The trends are smoother and clearer if the $\alpha$ parameter are fixed to the averaged value corresponding to the given centrality bin (for the averaged value, see Figure \ref{fig:alpha}.). 

\subsection{The L\'evy strength $\lambda$}

\begin{figure}
\centering
\includegraphics[width=0.45\textwidth]{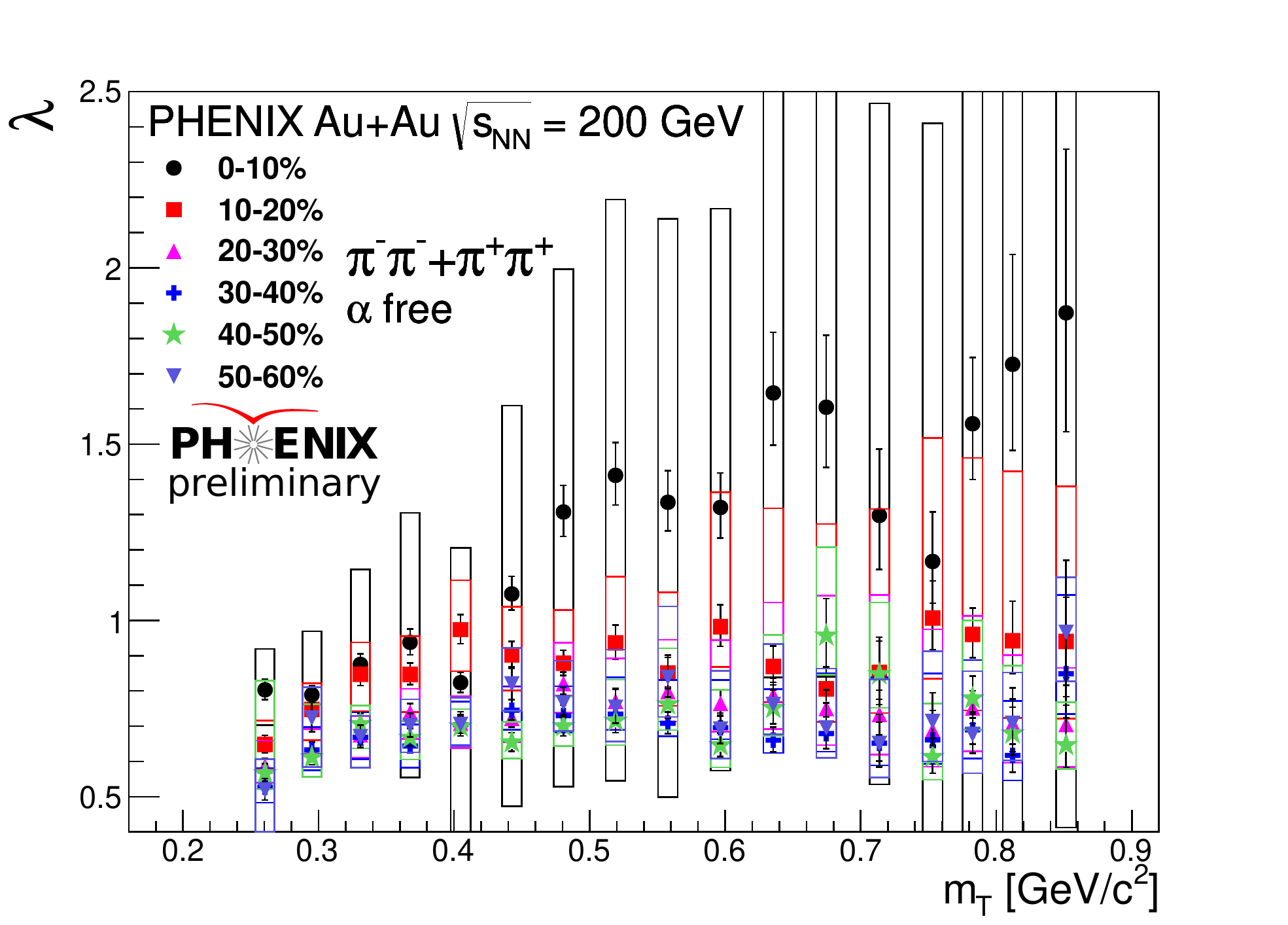}
\includegraphics[width=0.45\textwidth]{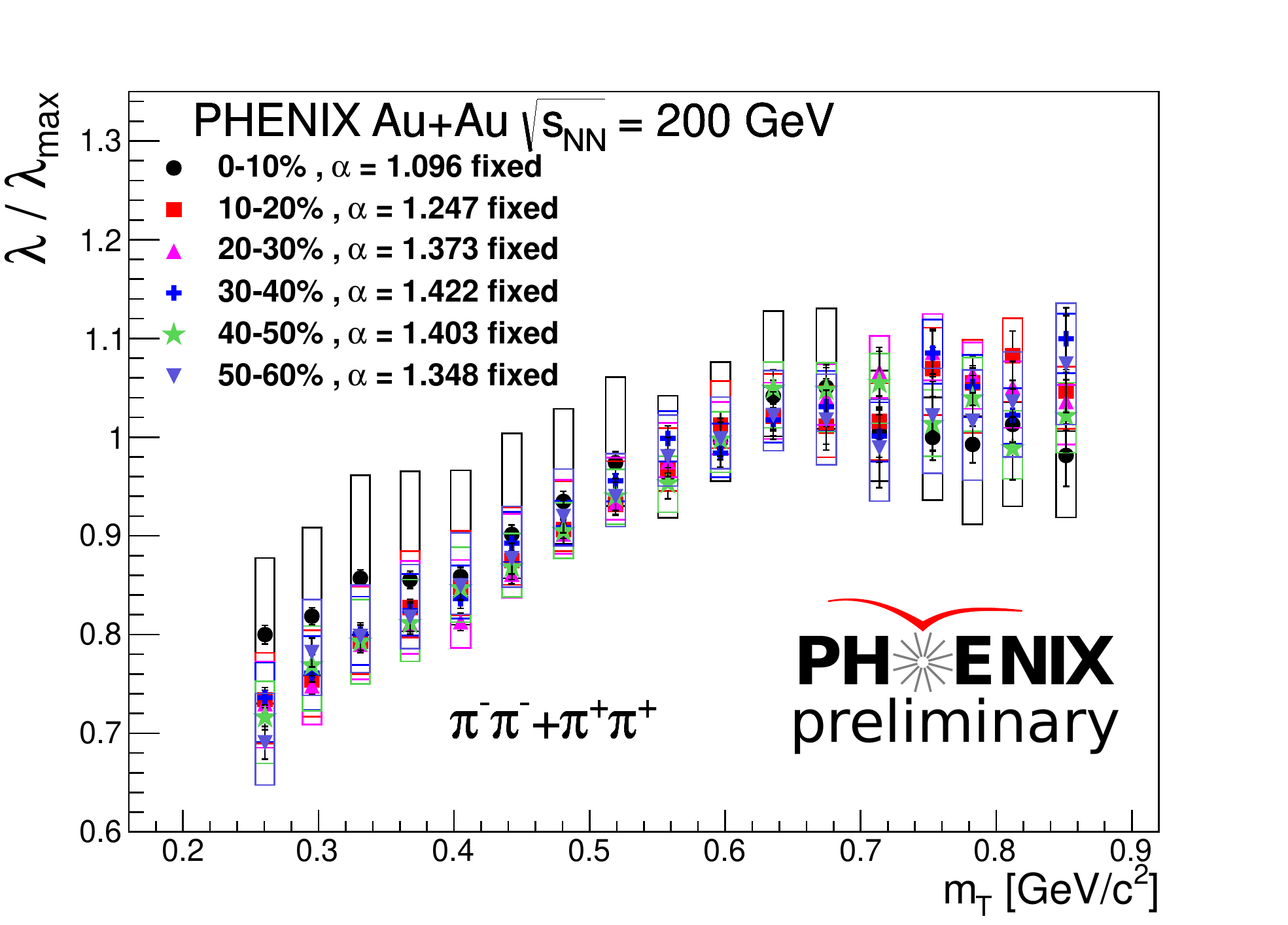}
\caption{\label{fig:lambda} The $\lambda(m_{\rm T})$ function in the different centrality ranges (left). The normalized $\lambda(m_{\rm T})$ function with fixed $\alpha=\langle \alpha \rangle$ value (right). The statistical errors are shown with errorbars and the systematical uncertainties boxes.}
\end{figure}

As it was mentioned above the $\lambda$ parameter could be related to the core-halo ratio. $\lambda(m_{\rm T})$ is shown in Figure \ref{fig:lambda}. The fixed $\alpha$ fit gives clearer and smoother trends. We introduced the normalized $\lambda$ function with a normalization factor $\lambda_{\rm max}= \langle \lambda \rangle_{(0.7-0.9)\textmd{GeV}/\textmd{c}^2}$. If we plot the $\lambda/\lambda_{\rm max}$ as a function of $m_{\rm T}$ it shows that the decreasing trends, or the characteristic of the ``hole'' does not depend strongly on the centrality.

\subsection{New scaling parameter}

\begin{figure}
\centering
\includegraphics[width=0.45\textwidth]{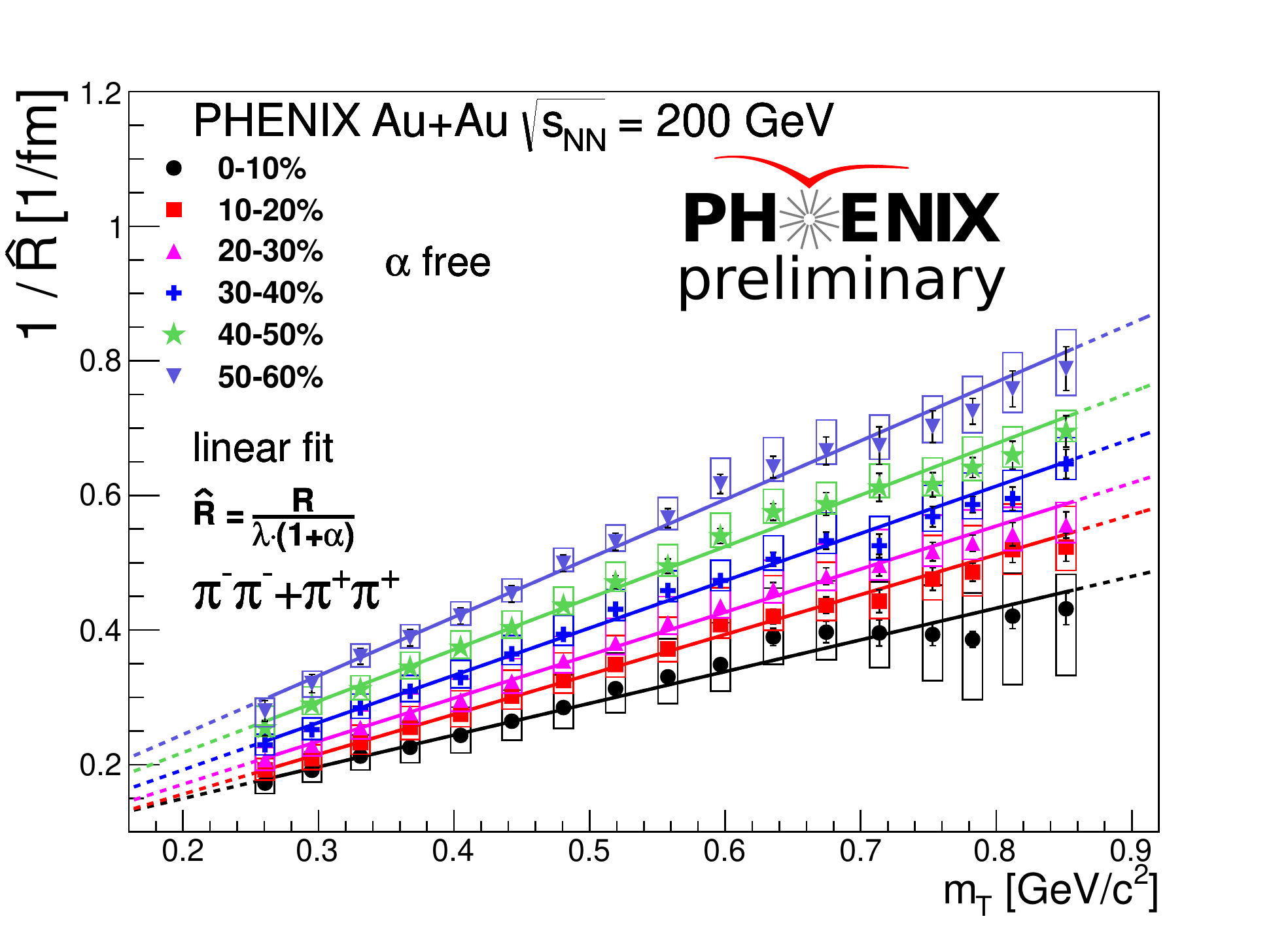}
\includegraphics[width=0.45\textwidth]{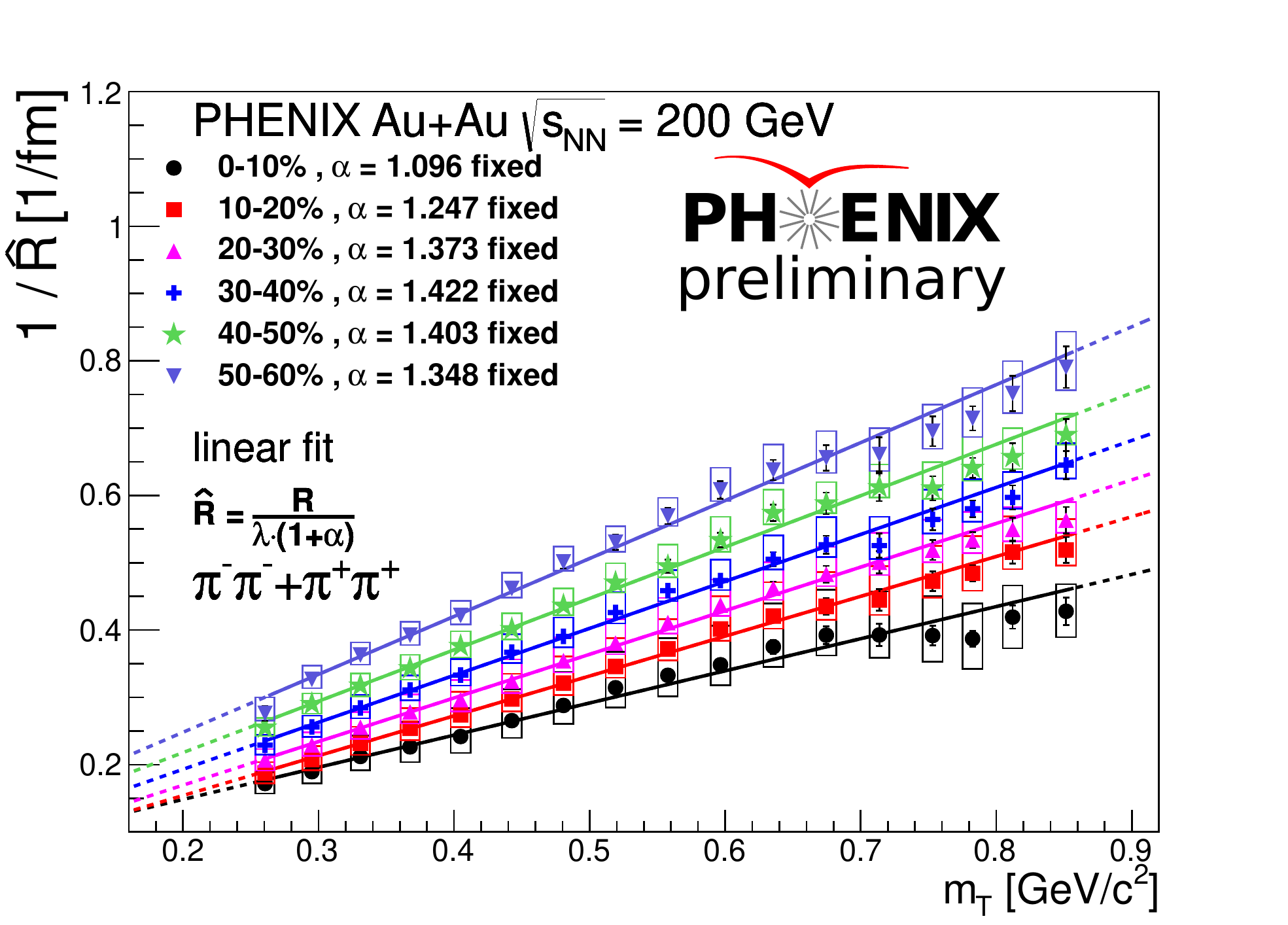}
\caption{\label{fig:rhat} The new, empirically found scaling variable does not depend on the $\alpha$ fixation. The statistical errors are shown with errorbars and the systematical uncertainties boxes.}
\end{figure}

The empirically found new scaling parameter \cite{Adare:2017vig, MateBGL} was observed in the centrality dependent case as well. If one introduces the combination of the three L\'evy parameters:

\begin{equation}
\hat{R} = \frac{R}{\lambda(1+\alpha)}
\end{equation}
a surprising scaling property of this variable rises up in all centrality ranges and in the whole investigated $m_{\rm T}$ range as it can be seen in Figure \ref{fig:rhat}.
A very remarkable property of this scaling is that it is not sensitive to the $\alpha$ fixation, even though all the parameters above get clearer and smoother trends with fixed $\alpha=\langle \alpha \rangle$ fits. In the case of this variable the $\alpha$ could be free or fixed, the trends are the same within statistical errors. The precise meaning of this variable is not clear yet and as we know there are not any theoretical prediction or explanation of it.

\section{Conclusion}

We measured correlation functions at 200 GeV at PHENIX experiment in 18 $m_{\rm T}$ and 6 centrality range. We fit the measured data with Coulomb corrected  L\'evy distribution, yield the L\'evy parameters and determine their $m_{\rm T}$ and centrality dependencies. We can find that the L\'evy-stability index $\alpha$ has non-monotonic behavior as a function of centrality. We also observed that the $\alpha$ is far from the Gaussian case corresponds to $\alpha=2$. The L\'evy scale parameter $R$ has hydrodynamical trends in all observed centrality bin as a function of $m_{\rm T}$. The strength of the correlation function $\lambda(m_{\rm T})$ decreases at lower $m_{\rm T}$. We also observed that the normalized $\lambda / \lambda_{\rm max}$ function has no strong centrality dependence. Further detailed and dedicated analysis is needed to clarify these observed phenomena.

\acknowledgments{The author is thankful for the support of EFOP-3.6.1-16-2016-00001 and NKFIH grant FK-123842}



\begin{thebibliography}{-------}
\providecommand{\natexlab}[1]{#1}

\bibitem[Brown and Twiss(1956)]{HBT_orig}
Brown, R.H.; Twiss, R.Q.
\newblock Correlation between Photons in two Coherent Beams of Light.
\newblock {\em Nature} {\bf 1956}, {\em 177},~27--29.

\bibitem[Goldhaber \em{et~al.}(1960)Goldhaber, Goldhaber, Lee, and
  Pais]{PhysRev.120.300}
Goldhaber, G.; Goldhaber, S.; Lee, W.; Pais, A.
\newblock Influence of Bose-Einstein Statistics on the Antiproton-Proton
  Annihilation Process.
\newblock {\em Phys. Rev.} {\bf 1960}, {\em 120},~300--312.

\bibitem[Adler \em{et~al.}(2007)Adler et~al.]{Adler:2006as}
Adler, S.S.; others.
\newblock {Evidence for a long-range component in the pion emission source in
  Au + Au collisions at s(NN)**(1/2) = 200-GeV}.
\newblock {\em Phys. Rev. Lett.} {\bf 2007}, {\em 98},~132301,
  \href{http://xxx.lanl.gov/abs/nucl-ex/0605032}{{\normalfont
  [arXiv:nucl-ex/nucl-ex/0605032]}}.

\bibitem[Cs{\"o}rg{\H o} \em{et~al.}(2004)Cs{\"o}rg{\H o}, Hegyi, and
  Zajc]{Csorgo:2003uv}
Cs{\"o}rg{\H o}, T.; Hegyi, S.; Zajc, W.A.
\newblock {Bose-Einstein correlations for Levy stable source distributions}.
\newblock {\em Eur. Phys. J.} {\bf 2004}, {\em C36},~67--78,
  \href{http://xxx.lanl.gov/abs/nucl-th/0310042}{{\normalfont
  [arXiv:nucl-th/nucl-th/0310042]}}.

\bibitem[Metzler \em{et~al.}(1999)Metzler, Barkai, and Klafter]{Metzler:1999zz}
Metzler, R.; Barkai, E.; Klafter, J.
\newblock {Anomalous Diffusion and Relaxation Close to Thermal Equilibrium: A
  Fractional Fokker-Planck Equation Approach}.
\newblock {\em Phys. Rev. Lett.} {\bf 1999}, {\em 82},~3563--3567.

\bibitem[Csan\'ad \em{et~al.}(2007)Csan\'ad, Cs{\"o}rg{\H o}, and
  Nagy]{Csanad:2007fr}
Csan\'ad, M.; Cs{\"o}rg{\H o}, T.; Nagy, M.
\newblock {Anomalous diffusion of pions at RHIC}.
\newblock {\em Braz. J. Phys.} {\bf 2007}, {\em 37},~1002--1013,
  \href{http://xxx.lanl.gov/abs/hep-ph/0702032}{{\normalfont
  [arXiv:hep-ph/hep-ph/0702032]}}.

\bibitem[Csan{\'a}d(2006)]{Csanad:2005nr}
Csan{\'a}d, M.
\newblock {Measurement and analysis of two- and three-particle correlations}.
\newblock {\em Nucl. Phys.} {\bf 2006}, {\em A774},~611--614,
  \href{http://xxx.lanl.gov/abs/nucl-ex/0509042}{{\normalfont
  [arXiv:nucl-ex/nucl-ex/0509042]}}.

\bibitem[Kincses(2016)]{Kincses:2016jsr}
Kincses, D.
\newblock {PHENIX results on L\'evy analysis of Bose-Einstein correlation
  functions}.
\newblock  {10th International Workshop on Critical Point and Onset of
  Deconfinement (CPOD 2016) Wrocław, Poland, May 30-June 4, 2016},  2016,
  \href{http://xxx.lanl.gov/abs/1610.05025}{{\normalfont
  [arXiv:nucl-ex/1610.05025]}}.

\bibitem[Adare \em{et~al.}(2017)Adare et~al.]{Adare:2017vig}
Adare, A.; others.
\newblock {L\'evy-stable two-pion Bose-Einstein correlations in
  $\sqrt{s_{_{NN}}}=200$ GeV Au$+$Au collisions} {\bf 2017}.
\newblock  \href{http://xxx.lanl.gov/abs/1709.05649}{{\normalfont
  [arXiv:nucl-ex/1709.05649]}}.

\bibitem[Csan\'ad()]{MateBGL}
Csan\'ad, M.
\newblock {Lévy femtoscopy with PHENIX at RHIC}.
\newblock  \href{http://xxx.lanl.gov/abs/1711.05575}{{\normalfont
  [1711.05575]}}.

\bibitem[Adcox \em{et~al.}(2003)Adcox et~al.]{Adcox:2003zm}
Adcox, K.; others.
\newblock {PHENIX detector overview}.
\newblock {\em Nucl. Instrum. Meth.} {\bf 2003}, {\em A499},~469--479.

\bibitem[Yano and Koonin(1978)]{Yano:1978gk}
Yano, F.B.; Koonin, S.E.
\newblock {Determining Pion Source Parameters in Relativistic Heavy Ion
  Collisions}.
\newblock {\em Phys. Lett.} {\bf 1978}, {\em 78B},~556--559.

\bibitem[Bolz \em{et~al.}(1993)Bolz, Ornik, Plumer, Schlei, and
  Weiner]{Bolz:1992hc}
Bolz, J.; Ornik, U.; Plumer, M.; Schlei, B.R.; Weiner, R.M.
\newblock {Resonance decays and partial coherence in Bose-Einstein
  correlations}.
\newblock {\em Phys. Rev.} {\bf 1993}, {\em D47},~3860--3870.

\bibitem[Cs{\"o}rg{\H o} \em{et~al.}(1996)Cs{\"o}rg{\H o}, L{\"o}rstad, and
  Zim{\'an}yi]{Csorgo:1994in}
Cs{\"o}rg{\H o}, T.; L{\"o}rstad, B.; Zim{\'an}yi, J.
\newblock {Bose-Einstein correlations for systems with large halo}.
\newblock {\em Z. Phys.} {\bf 1996}, {\em C71},~491--497,
  \href{http://xxx.lanl.gov/abs/hep-ph/9411307}{{\normalfont
  [arXiv:hep-ph/hep-ph/9411307]}}.

\bibitem[Cs\"org\H{o}(2002)]{Csorgo:1999sj}
Cs\"org\H{o}, T.
\newblock {Particle interferometry from 40-MeV to 40-TeV}.
\newblock {\em Heavy Ion Phys.} {\bf 2002}, {\em 15},~1--80,
  \href{http://xxx.lanl.gov/abs/hep-ph/0001233}{{\normalfont
  [arXiv:hep-ph/hep-ph/0001233]}}.

\bibitem[V\'ertesi \em{et~al.}(2011)V\'ertesi, Cs{\"o}rg{\H o}, and
  Sziklai]{Vertesi:2009wf}
V\'ertesi, R.; Cs{\"o}rg{\H o}, T.; Sziklai, J.
\newblock {Significant in-medium $\eta$' mass reduction in $\sqrt{s_{NN}}=200$
  GeV Au+Au collisions at the BNL Relativistic Heavy Ion Collider}.
\newblock {\em Phys. Rev.} {\bf 2011}, {\em C83},~054903,
  \href{http://xxx.lanl.gov/abs/0912.0258}{{\normalfont
  [arXiv:nucl-ex/0912.0258]}}.

\bibitem[Cs{\"o}rg{\H o} \em{et~al.}(2011)Cs{\"o}rg{\H o}, V{\'e}rtesi, and
  Sziklai]{Csorgo:2010hj}
Cs{\"o}rg{\H o}, T.; V{\'e}rtesi, R.; Sziklai, J.
\newblock {$U_A(1)$ Symmetry Restoration from an In-Medium $\eta$ ' Mass
  Reduction in $\sqrt{s_{NN}}=200$ GeV Au+Au Collisions}.
\newblock  {Quantum chromodynamics and beyond: Gribov-80 memorial volume.
  Proceedings, Memorial Workshop devoted to the 80th birthday of V.N. Gribov,
  Trieste, Italy, May 26-28, 2010},  2011, pp. 307--318,
  \href{http://xxx.lanl.gov/abs/1012.5058}{{\normalfont
  [arXiv:nucl-ex/1012.5058]}}.

\bibitem[Cs{\"o}rg{\H o} \em{et~al.}(2006)Cs{\"o}rg{\H o}, Hegyi, Nov{\'a}k,
  and Zajc]{Csorgo:2005it}
Cs{\"o}rg{\H o}, T.; Hegyi, S.; Nov{\'a}k, T.; Zajc, W.A.
\newblock {Bose-Einstein or HBT correlation signature of a second order QCD
  phase transition}.
\newblock {\em AIP Conf. Proc.} {\bf 2006}, {\em 828},~525--532,
  \href{http://xxx.lanl.gov/abs/nucl-th/0512060}{{\normalfont
  [arXiv:nucl-th/nucl-th/0512060]}}.

\bibitem[Gavrilik and Mishchenko(2015)]{Gavrilik:2014pxa}
Gavrilik, A.M.; Mishchenko, {\relax Yu}.A.
\newblock {Correlation function intercepts for $\tilde{\mu},q$-deformed Bose
  gas model implying effective accounting for interaction and compositeness of
  particles}.
\newblock {\em Nucl. Phys.} {\bf 2015}, {\em B891},~466--481,
  \href{http://xxx.lanl.gov/abs/1411.5955}{{\normalfont
  [arXiv:hep-ph/1411.5955]}}.

\bibitem[Sinyukov \em{et~al.}(1998)Sinyukov, Lednicky, Akkelin, Pluta, and
  Erazmus]{Sinyukov:1998fc}
Sinyukov, {\relax Yu}.; Lednicky, R.; Akkelin, S.V.; Pluta, J.; Erazmus, B.
\newblock {Coulomb corrections for interferometry analysis of expanding hadron
  systems}.
\newblock {\em Phys. Lett.} {\bf 1998}, {\em B432},~248--257.

\bibitem[Makhlin and Sinyukov(1988)]{Makhlin1988}
Makhlin, A.N.; Sinyukov, Y.M.
\newblock The hydrodynamics of hadron matter under a pion interferometric
  microscope.
\newblock {\em Zeitschrift f{\"u}r Physik C Particles and Fields} {\bf 1988},
  {\em 39},~69--73.

\bibitem[Cs{\"o}rg{\H o} and L{\"o}rstad(1996)]{Csorgo:1995bi}
Cs{\"o}rg{\H o}, T.; L{\"o}rstad, B.
\newblock {Bose-Einstein correlations for three-dimensionally expanding,
  cylindrically symmetric, finite systems}.
\newblock {\em Phys. Rev.} {\bf 1996}, {\em C54},~1390--1403,
  \href{http://xxx.lanl.gov/abs/hep-ph/9509213}{{\normalfont
  [arXiv:hep-ph/hep-ph/9509213]}}.

\bibitem[Chapman \em{et~al.}(1995)Chapman, Scotto, and Heinz]{Chapman:1994yv}
Chapman, S.; Scotto, P.; Heinz, U.W.
\newblock {A New cross term in the two particle HBT correlation function}.
\newblock {\em Phys. Rev. Lett.} {\bf 1995}, {\em 74},~4400--4403,
  \href{http://xxx.lanl.gov/abs/hep-ph/9408207}{{\normalfont
  [arXiv:hep-ph/hep-ph/9408207]}}.

\end{thebibliography}
\end{document}